\documentclass[12pt]{article}

\usepackage{epsfig}
\usepackage{latexsym}

\begin{document}
\noindent
DESY 02-121\hfill{\tt hep-lat/0208051\_v2}\\
March 2003
\vspace{5pt}

\begin{center}
{\Large\bf $M_\pi^2$ versus $m_q$: comparing CP-PACS and UKQCD data
to Chiral Perturbation Theory}
\end{center}
\vspace{5pt}

\begin{center}
{\bf\large Stephan D\"urr$\;{}^{a,b}$}\vspace{1mm}\\
{\small\it ${}^a\,$DESY Zeuthen, 15738 Zeuthen, Germany}\\
{\small\it ${}^b\,$INT at University of Washington, Seattle, WA 98195-1550,
U.S.A.}
\end{center}
\vspace{5pt}

\begin{abstract}
\noindent
I present a selection of CP-PACS and UKQCD data for the pseudo-Goldstone masses
in $N_f\!=\!2$ QCD with doubly degenerate quarks. At least the more chiral
points should be consistent with Chiral Perturbation Theory for the latter to
be useful in an extrapolation to physical masses. I find consistency with the
chiral prediction but no striking evidence for chiral logs. Nonetheless, the
consistency guarantees that the original estimate, by Gasser and Leutwyler, of
the $N_f\!=\!2$ QCD low-energy scale $\Lambda_3$ was not entirely wrong.
\end{abstract}


\newcommand{\pad}{\partial}
\newcommand{\pas}{\partial\!\!\!/}
\newcommand{\Dsl}{D\!\!\!\!/\,}
\newcommand{\Psl}{P\!\!\!\!/\;\!}
\newcommand{\hqu}{\hbar}
\newcommand{\ovr}{\over}
\newcommand{\til}{\tilde}
\newcommand{\pri}{^\prime}
\renewcommand{\dag}{^\dagger}
\newcommand{\<}{\langle}
\renewcommand{\>}{\rangle}
\newcommand{\gaf}{\gamma_5}
\newcommand{\lap}{\triangle}
\newcommand{\trc}{\rm tr}
\newcommand{\al}{\alpha}
\newcommand{\be}{\beta}
\newcommand{\ga}{\gamma}
\newcommand{\de}{\delta}
\newcommand{\ep}{\epsilon}
\newcommand{\ve}{\varepsilon}
\newcommand{\ze}{\zeta}
\newcommand{\et}{\eta}
\renewcommand{\th}{\theta}
\newcommand{\vt}{\vartheta}
\newcommand{\io}{\iota}
\newcommand{\ka}{\kappa}
\newcommand{\la}{\lambda}
\newcommand{\rh}{\rho}
\newcommand{\vr}{\varrho}
\newcommand{\si}{\sigma}
\newcommand{\ta}{\tau}
\newcommand{\ph}{\phi}
\newcommand{\vp}{\varphi}
\newcommand{\ch}{\chi}
\newcommand{\ps}{\psi}
\newcommand{\om}{\omega}
\newcommand{\psb}{\overline{\psi}}
\newcommand{\etb}{\overline{\eta}}
\newcommand{\psd}{\psi^{\dagger}}
\newcommand{\etd}{\eta^{\dagger}}
\newcommand{\beq}{\begin{equation}}
\newcommand{\eeq}{\end{equation}}
\newcommand{\bdm}{\begin{displaymath}}
\newcommand{\edm}{\end{displaymath}}
\newcommand{\bea}{\begin{eqnarray}}
\newcommand{\eea}{\end{eqnarray}}
\newcommand{\mr}{\mathrm}
\newcommand{\mb}{\mathbf}
\newcommand{\Nf}{N_{\!f\,}}
\newcommand{\Nc}{N_{\!c\,}}
\newcommand{\MeV}{\,\mr{MeV}}
\newcommand{\GeV}{\,\mr{GeV}}
\newcommand{\TeV}{\,\mr{TeV}}
\newcommand{\fm}{\,\mr{fm}}
\newcommand{\MSbar}{{\overline{\mr{MS}}}}
\newcommand{\alV}{\alpha_{{}_V}}
\newcommand{\alL}{\alpha_{{}_L}}


\hyphenation{topo-lo-gi-cal simu-la-tion theo-re-ti-cal mini-mum maxi-mum
re-nor-malize re-nor-maliza-tion}


\section*{Introduction}\vspace{-4pt}

Ever since numerical lattice QCD computations have been done, the spectrum of
light mesons has served as a benchmark problem.
This is still true today, as actions respecting the global chiral symmetry
among the light flavours are being tested and several groups have embarked on
ambitious simulations of ``full'' QCD with two or possibly more dynamical
flavours.
These developments address one of the key issues in low-energy QCD.
The fact that chiral symmetry is both spontaneously and explicitly broken
generates pseudo-Goldstone bosons, i.e.\ particles that dominate (for small
enough quark masses) the long-range behaviour of correlators between external
currents and which are collectively called ``pions''.

The lattice is not the only framework to address the low-energy structure of
QCD.
In the old days PCAC relations were exploited to predict the dependence of
low-energy observables on shifts in the quark masses and external momenta.
The one best known is
\beq
F_\pi^2 M_\pi^2=(m_\mr{u}\!+\!m_\mr{d})\;|\<0|\bar q q|0\>|+O(m^2)
\label{GOR}
\eeq
which connects the pion mass and decay constant to the product of the explicit
and spontaneous symmetry breaking parameters.
However, the Gell-Mann\,--\,Oakes\,--\,Renner relation (\ref{GOR}) does not
give a prediction how $F_\pi$ and $M_\pi$ separately depend on the quark mass.
And in the real world the latter may be shifted only by a discrete amount
(e.g.\ by replacing $d\to s, \pi\to K$ one gets a leading order prediction for
the quark mass ratio $m_s/m$ with $m\equiv(m_u\!+\!m_d)/2$).

Today, the first limitation is overcome, since the successor of PCAC, Chiral
Perturbation Theory \cite{Gasser:1983yg}, gives detailed predictions how either
$M_\pi$ or $F_\pi$ individually depends on the quark mass $m$.
And --~in principle~-- the second problem is gone since one may vary $m$
continuously in a lattice computation.
Therefore, it seems natural to combine the two approaches to benefit from their
respective advantages.
However, for that aim quarks need to be taken sufficiently light, and this is a
numerical challenge on the lattice.
Below, an elementary test is presented whether this is already achieved in
present state-of-the-art results for $M_\pi^2$ versus $m$, as published by the
CP-PACS \cite{AliKhan:2001tx} and UKQCD \cite{Allton:2001sk} collaborations.
The idea is to restrict the analysis to the ``doubly degenerate'' case with
$\Nf\!=\!2$ $O(a)$ improved Wilson quarks, i.e.\ to consider only the subset
where both valence-quarks are exactly degenerate with the sea-quarks, as this
avoids additional assumptions of the ``partially quenched'' framework.

\clearpage


\section*{Prediction by Chiral Perturbation Theory}

Here, I give a brief summary of the work by Gasser and Leutwyler (GL), with an
application to lattice data in mind.

GL have calculated the substitute for the GOR relation (\ref{GOR}) to NLO in
Chiral Perturbation Theory (XPT) with $\Nf\!=\!2$ quarks \cite{Gasser:1983yg}.
To that order, two low-energy constants%
\footnote{Unlike $F$, the other LO ``constant'' $B$ is scheme- and
scale-dependent, as is $m$. In the chiral representation, only the product $mB$
appears which is RG-invariant (cf.\ (\ref{Msqdef})). In order to
determine $B$ and $m$ separately, a lattice computation is needed; the result
of the CP-PACS study \cite{AliKhan:2001tx} along with (\ref{Msqdef}) for the
physical pion is
\beq
m_\mr{phys}(\MSbar,\mu\!\sim\!2\GeV)\simeq3.5\pm0.2\MeV\;,\quad
B(\MSbar,\mu\!\sim\!2\GeV)\simeq2.8\mp0.15\GeV\;.
\label{Besti}
\eeq}
from the LO Lagrangian appear ($F\!=\!\lim_{m\to0}F_\pi,
B\!=\!-\lim_{m\to0}\<0|\bar q q|0\>/F^2$) together with the finite parts of the
$l_i$ in the NLO Lagrangian.
Restricted to the degenerate case their result \cite{Gasser:1983yg} takes the
form
\bea
M_\pi^2\!&\!=\!&\!M^2+
{1\ovr F^2}\Big({M^4\ovr32\pi^2}\log({M^2\ovr\mu^2})-2M^4\,l_3^\mr{r}(\mu)\Big)
\label{Mpiori}
\\
F_\pi\!&\!=\!&\!F+
{1\ovr F}\Big(-\!{M^2\ovr16\pi^2}\log({M^2\ovr\mu^2})+M^2\,l_4^\mr{r}(\mu)\Big)
\label{Fpiori}
\\
M^2\!&\!\equiv\!&\!2mB\;.
\label{Msqdef}
\eea
Note that (\ref{Mpiori}, \ref{Fpiori}) has the typical structure of a NLO
prediction: A {\em chiral logarithm\/} summarizing the contribution from the
pion loops appears together with a {\em counterterm\/}.
The $l_i^\mr{r}$ are the renormalized GL coefficients, i.e.\ they are the
descendents of the $l_i$ which appear in the NLO Lagrangian and which are
divergent quantities.
As a result, the $l_i^\mr{r}$ depend on the (chiral) renormalization scale
$\mu$. In the dimensionally regularized theory one has \cite{Gasser:1983yg}
\bea
l_i\!&\!=\!&\!l_i^\mr{r}(\mu)+\ga_i\la(\mu)
\\
\la(\mu)\!&\!=\!&\!{-1\ovr16\pi^2\mu^{4-d}}
\Big({1\ovr 4-d}+{\log(4\pi)+\Gamma\pri(1)+1\ovr2}\Big)
\\
l_i^\mr{r}(\mu)\!&\!=\!&\!l_i^\mr{r}(\mu^\star)-
{\ga_i\ovr16\pi^2}\log({\mu\ovr\mu^\star})
\eea
and the $\be$-function coefficients $\ga_i$ are known.
In the present context only $\ga_3=-{1\ovr2}\,,\;\ga_4=2$  \cite{Gasser:1983yg}
are relevant, and this suggests that one rewrites (\ref{Mpiori}, \ref{Fpiori})
with the help of
\beq
l_i^\mr{r}={\ga_i\ovr32\pi^2}\Big(\bar l_i+\log({M^2\ovr\mu^2})\Big)
\;,
\eeq
where the $\mu$ dependence in $l_i^\mr{r}\!=\!l_i^\mr{r}(\mu)$ is traded for
an $M$ dependence in $\bar l_i\!=\!\bar l_i(M)$, to get \cite{Gasser:1983yg}
\bea
M_\pi^2\!&\!=\!&\!M^2\,\Big(1-{M^2\ovr32\pi^2 F^2}\bar l_3 +O(M^4)\Big)
\label{Mpiint}
\\
F_\pi\!&\!=\!&\!F\;\Big(1+{M^2\ovr16\pi^2 F^2}\bar l_4 +O(M^4)\Big)
\;.
\label{Fpiint}
\eea
The NLO part is given in terms of the LO parameters $F, B$ and the NLO
coefficients $\bar l_3, \bar l_4$.
While the former two are known quite accurately, for the $\bar l_i$ only their
running in $M^2$ is known exactly and the phenomenological estimate of the
integration constants has comparatively large error-bars.
Gasser and Leutwyler give in their initial paper \cite{Gasser:1983yg} the
estimate
\beq
\bar l_3(m_\mr{phys})=2.9\pm2.4\;,\qquad\bar l_4(m_\mr{phys})=4.3\pm0.9
\label{l34esti}
\eeq
for real world quark masses.
It turns out that even this limited information is useful, since it determines
the {\em curvature\/} of $M_\pi$ as a function of the quark mass.
Close to the chiral limit, both $\bar l_i$ are {\em positive\/} and as a
consequence $M_\pi^2$ does not rise strictly linear in $m$ but turns
{\em right\/}, while $F_\pi$ has a {\em positive\/} first derivative in $m$.
More specifically, (\ref{l34esti}) translates into
\beq
M_{\pi,\mr{phys}}\!\simeq\!139\MeV\,,\;F_{\pi,\mr{phys}}\!\simeq\!92.4\MeV
\;\Longleftrightarrow\;
M_\mr{phys}\!\simeq\!141\MeV\,,\;F\!\simeq\!86.1\MeV
\label{Festi}
\eeq
which means that the physical pion is somewhat lighter than it would be, if the
LO relation were exact, while is decay constant exceeds its value in the
chiral limit by $\sim\!7\%$.

A seemingly formal point which, in the end, proves convenient in analyzing
the lattice data is the following.
A naive look at (\ref{Mpiint}) suggests that the typical structure of a NLO
prediction is gone -- rather than a $M^4$ and a $M^4\log(M^2)$
contribution, only the polynomial part seems left.
The point is that this impression is entirely misleading; the IR divergencies
(which are genuine to QCD in the chiral limit) are not gone, they are just
hidden in the $\bar l_i$.
The situation is, in fact, opposite -- the $M^4$ part has been eliminated in
favour of a pure $M^4\log(M^2)$ contribution, and the last step is to make this
apparent.
The quark mass dependence of $\bar l_3$ is given through
\beq
\log({\Lambda_i^2\ovr M^2})=
\log({\Lambda_i^2\ovr M_\mr{phys}^2})+
\log({M_\mr{phys}^2\ovr M^2})=
\bar l_i(m_\mr{phys})+
\log({M_\mr{phys}^2\ovr M^2})=
\bar l_i(m)
\label{Lambdai}
\eeq
and together with (\ref{Msqdef}), the relation takes its final form (see e.g.\
\cite{Leutwyler:2000hx})
\bea
M_\pi^2\!&\!=\!&\!2mB-{m^2B^2\ovr8\pi^2F^2}\log({\Lambda_3^2\ovr 2mB})+O(m^3)
\label{Mpiult}
\\
F_\pi\!&\!=\!&\!F+{mB\ovr8\pi^2F}\log({\Lambda_4^2\ovr 2mB})+O(m^3)
\label{Fpiult}
\eea
where $\Lambda_4, \Lambda_3$ represent {\em universal low energy scales\/}.
The estimates (\ref{l34esti}) translate into
\beq
\Lambda_3=0.6\GeV\begin{tabular}{l}$+1.4\GeV$\\$-0.4\GeV$\end{tabular}\;,\qquad
\Lambda_4=1.2\GeV\begin{tabular}{l}$+0.7\GeV$\\$-0.4\GeV$\end{tabular}
\;.
\label{L34esti}
\eeq
It is worth emphasizing that all four parameters $F, M^2\!=\!2Bm, \Lambda_3,
\Lambda_4$ do not depend on the QCD renormalization scheme \cite{Gasser:1983yg},
i.e.\ they are the proper physical low-energy parameters at order $O(p^4)$.
Furthermore the $\Lambda_i$ do not depend on the quark masses, and this means
that the representation (\ref{Mpiult}, \ref{Fpiult}) is perfectly suited to
analyze the lattice data even if they are gotten at quark masses larger than
$m\equiv(m_\mr{u}\!+\!m_\mr{d})/2$ in the real world -- as long as they are not
beyond the regime of validity of the chiral expansion itself.

The latter point is one of the key issues in the comparison we aim at.
The chiral expansion is known to be asymptotic, and this means that increasing
the order will enhance the accuracy near the chiral limit -- at the price of
worsening the prediction for heavier masses.
What is the ``critical scale'' beyond which the chiral expansion ``explodes''
is, a priori, not known.
From a formal point of view, one might think that (\ref{Mpiint}, \ref{Fpiint})
indicate that the expansion is in $M/(4\pi F)$ and hence hope that it is good
for pion masses up to $1\GeV$.
In this paper I will argue that watching the convergence pattern at a fixed
quark mass gives a more reliable estimate what is the permissible range.
This is facilitated since the NNLO expression for $M_\pi^2$ (with $\Nf\!=\!2$)
is known \cite{Colangelo:1995jm, Burgi:1996qi}.
The result reads (for the presentation I follow \cite{Leutwyler:2000hx})
\beq
M_\pi^2=2mB
\bigg(
1-{mB\ovr16\pi^2F^2}\log({\Lambda_3^2\ovr M^2})+{m^2B^2\ovr64\pi^4F^4}
\Big\{{17\ovr8}\Big(\log({\Lambda_M^2\ovr M^2})\Big)^2+k_M\Big\}+O(m^3)
\bigg)
\label{NNLO}
\eeq
where $\Lambda_M$ is implicitly defined through
$51\log({\Lambda_M^2\ovr\mu^2})=28\log({\Lambda_1^2\ovr\mu^2})+
32\log({\Lambda_2^2\ovr\mu^2})-9\log({\Lambda_3^2\ovr\mu^2})+49$
and the mass-independent $k_M$ accounts for the remainder at $O(p^6)$, in
particular the new counterterms.
Phenomenological values for $\Lambda_M$ and $k_M$ will be mentioned below.


\begin{figure}
\vspace{-4mm}
\epsfig{file=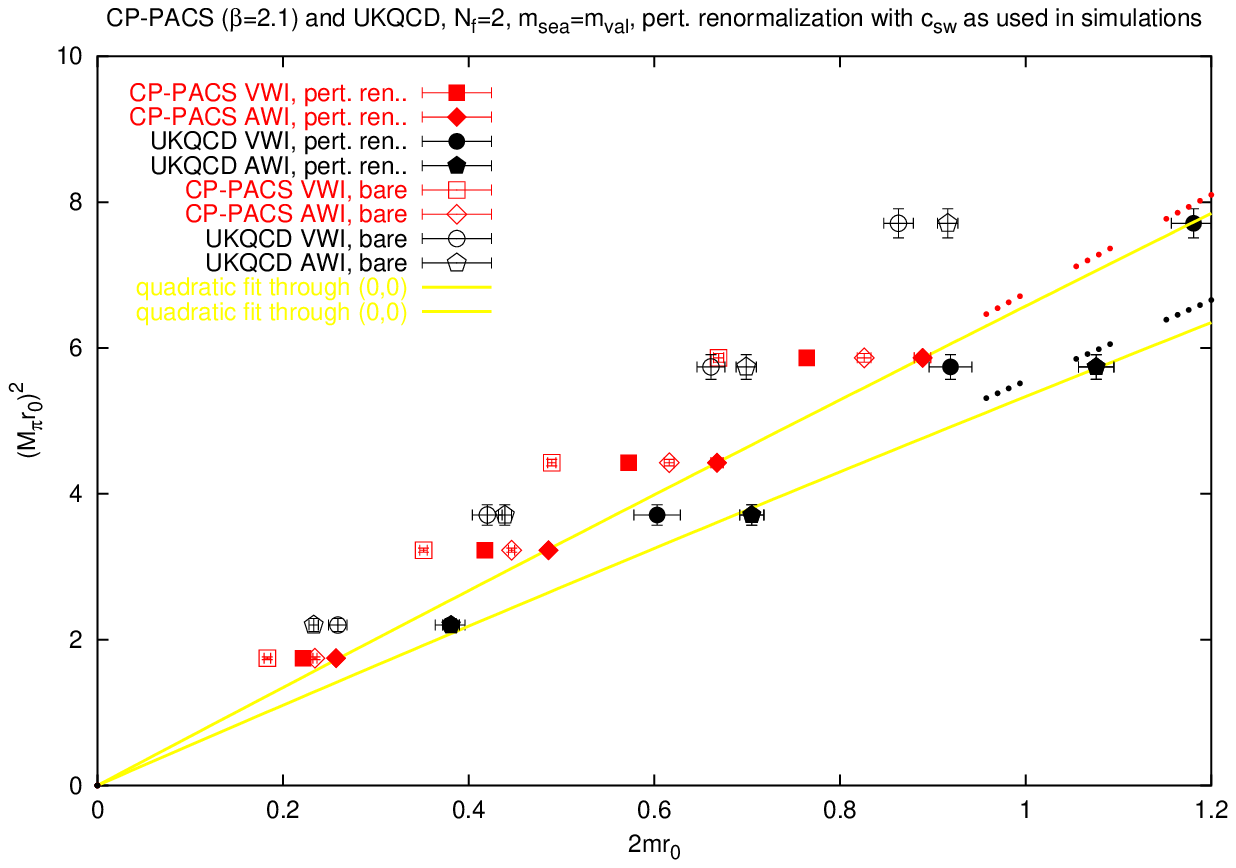,width=16.75cm,height=12cm,angle=0}
\vspace{-2mm}
\\
\epsfig{file=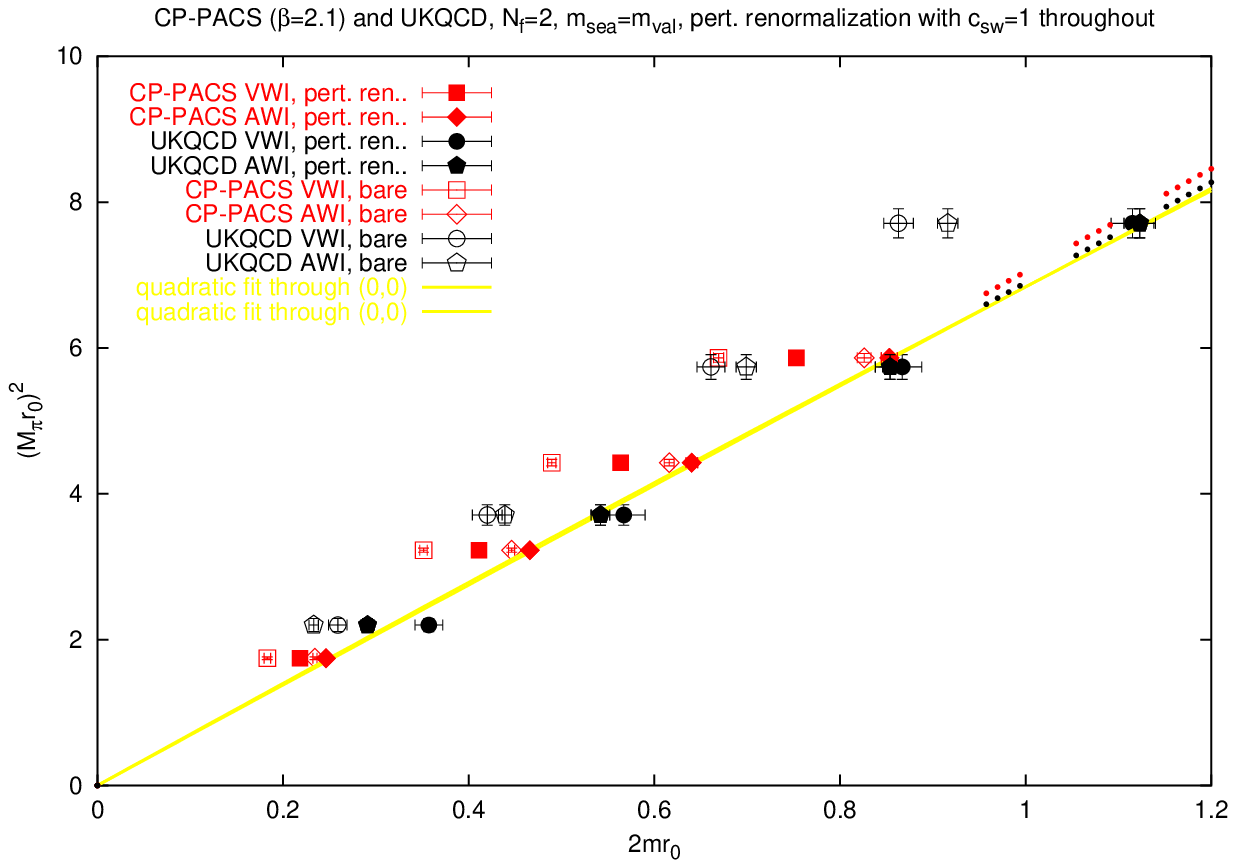,width=16.75cm,height=12cm,angle=0}
\vspace{-4mm}
\caption{\small\sl CP-PACS and UKQCD data converted to the form $M_\pi^2$ vs.\
$2m$ with masses in units of $r_0^{-1}$. Top: Perturbative renormalization with
$c_\mr{SW}$ as used in the simulations. Bottom: Same but $c_\mr{SW}=1$
throughout. A quadratic fit, constrained to go through zero, is applied to the
renormalized AWI data. Segments of the asymptotic slope in the chiral limit
highlight the curvature.}
\vspace{-1.5mm}
\end{figure}

\section*{Lattice Data}

We are now in a position to esteem the results by the CP-PACS and UKQCD
collaborations for the quark mass dependence of the pion mass.
We shall consider, out of these $\Nf\!=\!2$ data, only the two-fold degenerate
case where both valence quarks have the same mass and where they are, at the
same time, exactly degenerate with the sea quarks so that the theory is
unitary.

\bigskip

The CP-PACS collaboration has simulated various ($\be, \ka$) combinations with
an RG improved gauge action and a mean-field improved clover quark action
\cite{AliKhan:2001tx}.
They use a grid of size $12^3\!\times\!24, 16^3\!\times\!32, 24^3\!\times\!48,
24^3\!\times\!48$ at $\be\!=\!1.8, 1.95, 2.1, 2.2$, respectively, which leads
to a lattice spacing, if determined through the $\rh$ mass, between
$0.215\fm$ and $0.087\fm$ and hence a spatial box size between $2.58\fm$ and
$2.08\fm$.
With so much information at hand, one could, in principle, attempt a continuum
extrapolation for $M_\pi^2$ versus the sum of the (degenerate) quark masses,
$2m$.
However, non-perturbative renormalization might be necessary, and/or the
lattice spacing might be too large at the lower $\be$ values.
For this reason, I have decided to concentrate on the $\be\!=\!2.1$ data,
since here discretization effects are not supposed to be too large, and
good statistics is available.
At that $\be$ value, even the lightest pion is unlikely to suffer from
finite-size effects, since $M_\pi L\!>\!7$, and the lattice spacing determined
via the $\rh$ mass is of order $0.1\fm$ and hence comparable with that in the
UKQCD simulations.

The UKQCD collaboration works on a $16^3\times 32$ grid, using different
actions: Wilson glue and non-perturbatively $O(a)$ improved clover quarks
\cite{Allton:2001sk}.
Another point in which they differ from CP-PACS is a tactical one: they try to
relax $\be$ in pushing $\ka_\mr{sea}$ up (i.e.\ the quark mass down) such that
the lattice spacing, in units of $r_0$, stays constant.
Numerically, it is $a\!\sim\!0.1\,\mr{fm}$ for the data considered below
(though the ensemble at $(\be,\ka_\mr{sea})\!=\!(5.2,0.1355)$ is not matched
any more), and the hope is that the size of discretization effects would be
approximately constant.
This choice implies that the physical box size stays constant, too, and the
bound $M_\pi L\!>\!4.5$ maintained makes one feel comfortable that finite size
effects are small.

\bigskip

The plan of this article is to ignore that the dynamical quark masses might be
too heavy for the chiral prediction to be applicable and to ignore that in
principle a continuum extrapolation is needed but instead to go ahead and
simply compare the two datasets on a ``as-is'' basis with the LO/NLO/NNLO
prediction from Chiral Perturbation Theory in the continuum.
With the relevant low-energy constants on the chiral side given in physical
units, $r_0^{-1}$ must be so, too.
In this article, this is done through the assumption that $r_0^{-1}$ represents
a universal low-energy scale, unaffected by unquenching effects; the numerical
value used is $r_0\!=\!0.5\fm$ \cite{Sommer:1993ce}.

\bigskip

On the lattice, there are two definitions of the ``quark mass'', one through
the vector Ward-Takahashi identity (VWI mass), one through the axial identity
(AWI mass).
The bare masses (open symbols in Fig.\ 1) need not agree, while after
renormalization (filled symbols in Fig.\ 1) they would (up to $O(a^2)$ effects)
if the renormalization factors were computed non-perturbatively.
For unquenched ($\Nf\!=\!2$) data the relevant non-perturbative factors are not
yet available. 
For this reason, I have decided to renormalize both the CP-PACS and the UKQCD
data at one-loop order (the details being given in the appendices).
In the case of the CP-PACS data this basically repeats their calculation
\cite{AliKhan:2001tx}, albeit with two notable differences:
First, the scale is set through the measured $r_0$, since this is the only
possibility with the UKQCD data.
Second, the ``boost factor'' $u_0$ is derived from the measured plaquette
rather than the plaquette in the chiral limit, since in the UKQCD data there is
no uniquely defined chirally extrapolated version.
This perturbative calculation then lays the ground on which the CP-PACS and
UKQCD data may be compared to each other and to the chiral prediction.

Fig.\ 1 displays $(M_\pi r_0)^2$ versus the renormalized quark masses with
filled symbols, open symbols indicate the bare data to visualize the shift.
In the upper part the renormalization was performed with the $c_\mr{SW}$
values as they were used in the simulations (suggested by a mean-field analysis
\cite{AliKhan:2001tx} or the Alpha study \cite{Allton:2001sk, Jansen:1998mx}).
In the lower part the renormalization factors were computed with
$c_\mr{SW}\!=\!1$ throughout, which is a consistent choice at one-loop order.
Obviously, the overall consistency of the data is much better with this latter
choice, as is particularly obvious from the quadratic fits (constrained to go
through zero) to the AWI data.
Henceforth, we shall stick to the latter choice ($c_\mr{SW}\!=\!1$), but it is
useful to keep in mind that the difference to the upper part is a clear
indication of the size of inherent perturbative uncertainties.
Finally, it is worth mentioning that all error-bars in this article represent
only statistical errors.

\bigskip

We now turn to the physics content.
To convey a feeling for the scales, I mention that at $2mr_0\!=\!1$ the sum of
the valence quark masses is of order $400\MeV$, i.e.\ about four times as much
as in a physical $K$, and the corresponding ``pion'' weighs about $1\GeV$.
The physical kaon weighs $496\MeV$, i.e.\ $(M_K r_0)^2\!\simeq\!1.58$.
If the pion would satisfy $(M_\pi r_0)^2\!\simeq\!1.58$ this would mean that
its $u$- and $d$-quarks are about half as heavy as the $s$-quark in the real
world.
The lightest pion in the CP-PACS and UKQCD simulations have $(M_\pi r_0)^2$
values 1.75 and 2.20, respectively, and from this we conclude that their
lightest $u$- and $d$- quarks (in the unitary theory) have about 55\% and
70\% of the mass of the physical strange quark.

\begin{figure}[!t]
\vspace{-4mm}
\epsfig{file=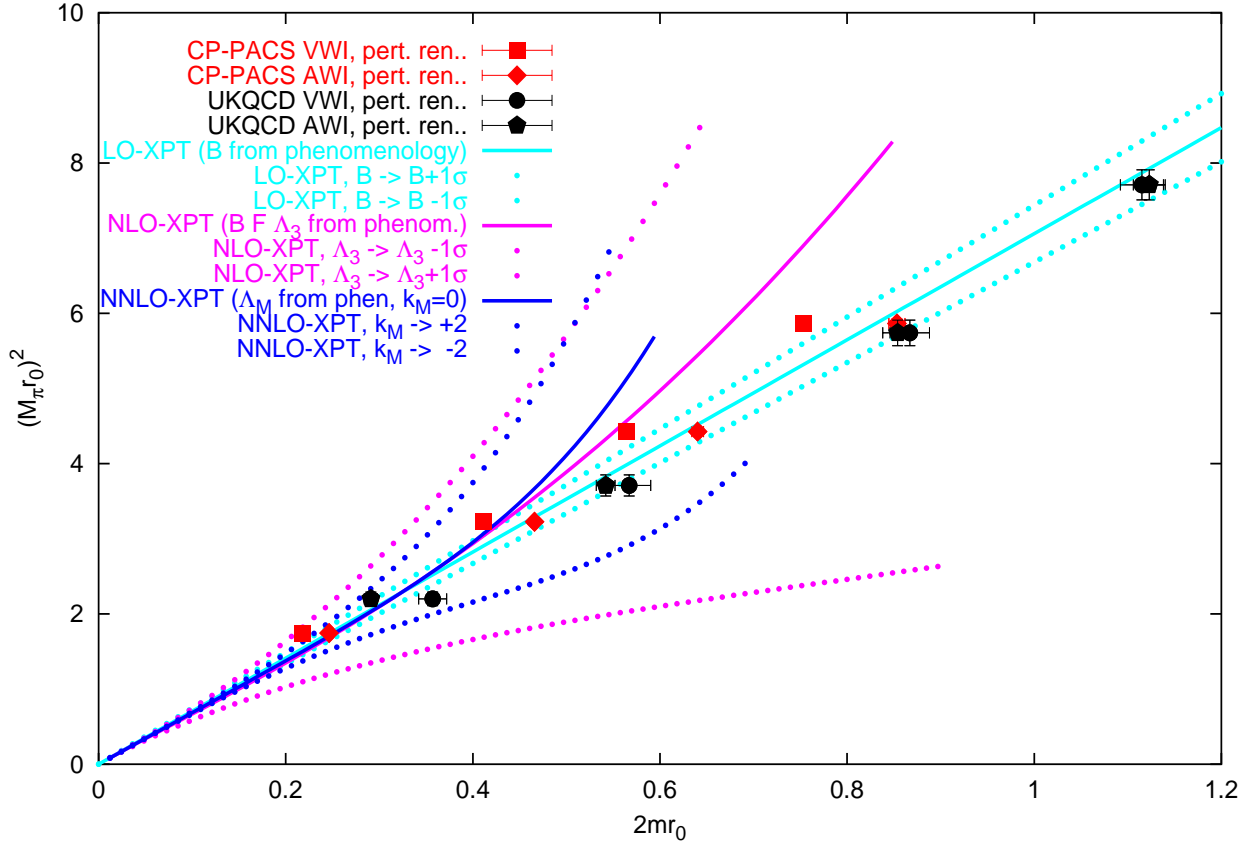,width=16.75cm,height=12cm,angle=0}
\vspace{-4mm}
\caption{\small\sl LO/NLO/NNLO chiral predictions with phenomenological values
for $B, F, \Lambda_3,\Lambda_M, k_M$ $(\pm1\si$ variation included at each
order, cf.\ text for details$)$ compared to the renormalized data in the
version with $c_\mr{SW}=1$. Note that the lines represent parameter-free
predictions, not fits.}
\end{figure}

With such heavy ``light'' quarks it is a priori not clear whether XPT is of any
use to extrapolate them to the physical $u$- and $d$-quark masses.
In an attempt to shed a light on this issue, Fig.\ 2 shows the renormalized
data (as in the bottom part of Fig.\ 1) together with the predictions from XPT
at tree/one-loop/two-loop level (LO/NLO/NNLO).
The low-energy constants are taken from phenomenology, i.e.\ these curves
represent {\em parameter-free predictions\/}.
At LO the chiral prediction is a straight line with the slope parameter $B$
taken from (\ref{Besti}); the $\pm1\si$ bounds are indicated by dotted lines.
At NLO the prediction is curved, and the numerical values of the additional
parameters are taken from (\ref{Festi}) and (\ref{L34esti}).
$\Lambda_3$ is varied within its phenomenological $\pm1\si$ bound (full versus
dotted lines -- lowering $\Lambda_3$ to $200\MeV$ yields the {\em upper\/},
increasing it to $2\GeV$ yields the {\em lower\/} dotted line), with $B, F$
fixed at their central values.
At NNLO the parameters $\Lambda_M, k_M$ in (\ref{NNLO}) are determined as
follows.
With $\Lambda_1\!=\!0.11\!\pm\!{0.04\atop0.03}\GeV,
\Lambda_2\!=\!1.2\!\pm\!0.06\GeV$ \cite{Colangelo:2001df} and (\ref{L34esti})
the relation beneath eqn.\ (\ref{NNLO}) gives
\beq
\Lambda_M=0.60\!\pm\!0.03\GeV
\label{LMesti}
\eeq
where errors have been added in quadrature.
This means that over the range considered the uncertainty in $\Lambda_M$ is
neglibigle compared to that coming from $k_M$.
Phenomenological arguments indicate $|k_M|\!\sim\!2$ \cite{LeutwylerPrivate},
and the sum-rule estimates for the NNLO counterterms given in
\cite{Colangelo:2001df} may be converted into a more accurate estimate.
For the purpose of the present article it is sufficient to use $k_M\!=\!0\pm2$,
and the associate curves (holding $B, F, \Lambda_3, \Lambda_M$ fixed at their
central values) are included in Fig.\ 2.

\begin{figure}
\vspace{-4mm}
\epsfig{file=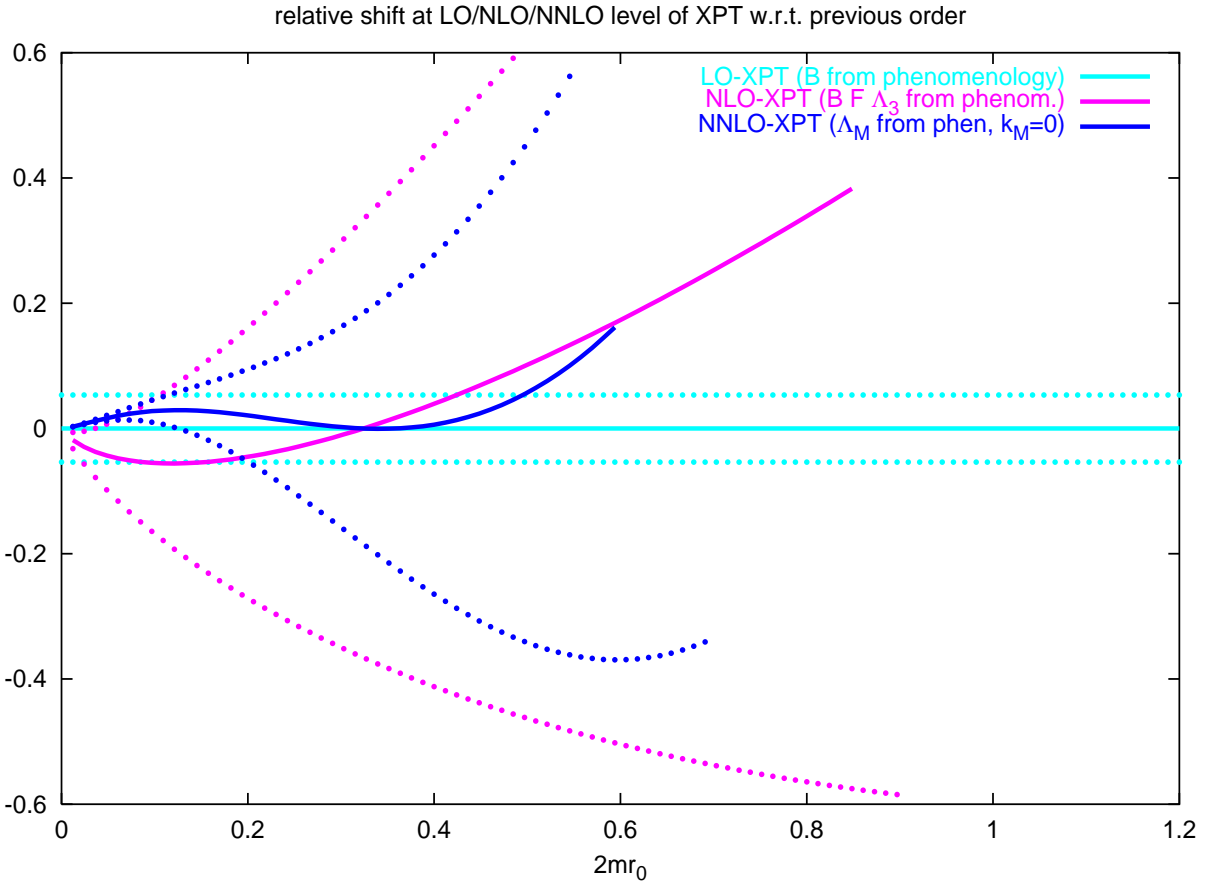,width=16.75cm,height=12cm,angle=0}
\vspace{-4mm}
\caption{\small\sl Relative shifts in $M_\pi^2$ versus $2mr_0$ in XPT -- due to
the uncertainty in $B$ at LO (light), due to NLO contribution with $B$ fixed at
central value, $\Lambda_3$ varied within $1\si$ bound (intermediate), and due
to NNLO contribution with $B$ and $\Lambda_3$ fixed at central values,
$k_M\in\{0,\pm 2\}$ (dark).}
\end{figure}

This information is sufficient to asses the {\em chiral convergence
behaviour\/} in $M_\pi^2$ versus $2m$.
To that aim we compare the LO/NLO/NNLO predictions for a given quark mass.
Fig.\ 3 shows the relative shift in $M_\pi^2$ when one more order is
included or the new low-energy constants are varied within reasonable bounds.
At first sight, the uncertainties at higher orders due to the error bars of the
counterterms seem large, but one should keep in mind that the associate shifts
are 100\% correlated over the whole range.
For instance, if the CP-PACS VWI point at $2mr_0\!\sim\!0.2$ sits on the
$-1\si$ curve (upper dotted NLO line in Fig.~2), then the one at
$2mr_0\!\sim\!0.4$ should too, if only $\Lambda_3$ needs to be adapted.
This means that precise lattice data are ideally suited to reduce the error on
$l_3^\mr{r}(\mu)$ (or $\Lambda_3$).
Fig.\ 3 shows that the crossover where the uncertainty due to the NLO exceeds
that due to the LO contribution is at $2mr_0=0.45/0.02/0.10$ when $\Lambda_3$
is taken at its central/+1$\sigma$/-1$\sigma$ value.
By averaging with weights 2/1/1 one arrives at the estimate that in the case of
$M_\pi$ the chiral expansion is sufficiently well-behaved that the NLO
functional form is useful for quark masses up to
\beq
2mr_0\leq0.25
\qquad \Longleftrightarrow \qquad (M_\pi r_0)^2\leq2
\qquad \Longleftrightarrow \qquad M_\pi\leq560\MeV
\label{range}
\;,
\eeq
and maybe more.
This, if correct, means that current state-of-the-art simulations make contact
with the regime where NLO-XPT holds, but so far a non-trivial ``{\em lever
arm\/}'' which is needed to make model independent predictions in the deeply
chiral regime, is likely {\em missing\/}.

\bigskip

While it is clear that future simulation data will allow to test the prediction
(\ref{range}) and provide, if it is correct, the lever arm needed, one might,
already at this time, go ahead and try what comes out if one assumes that the
estimate (\ref{range}) is too pessimistic and hence uses the NLO chiral ansatz
to fit the data over an extended range.
This is what we shall do below, but it is clear that this attempt is rather
speculative and results should be taken with care.

\begin{figure}
\vspace{-4mm}
\epsfig{file=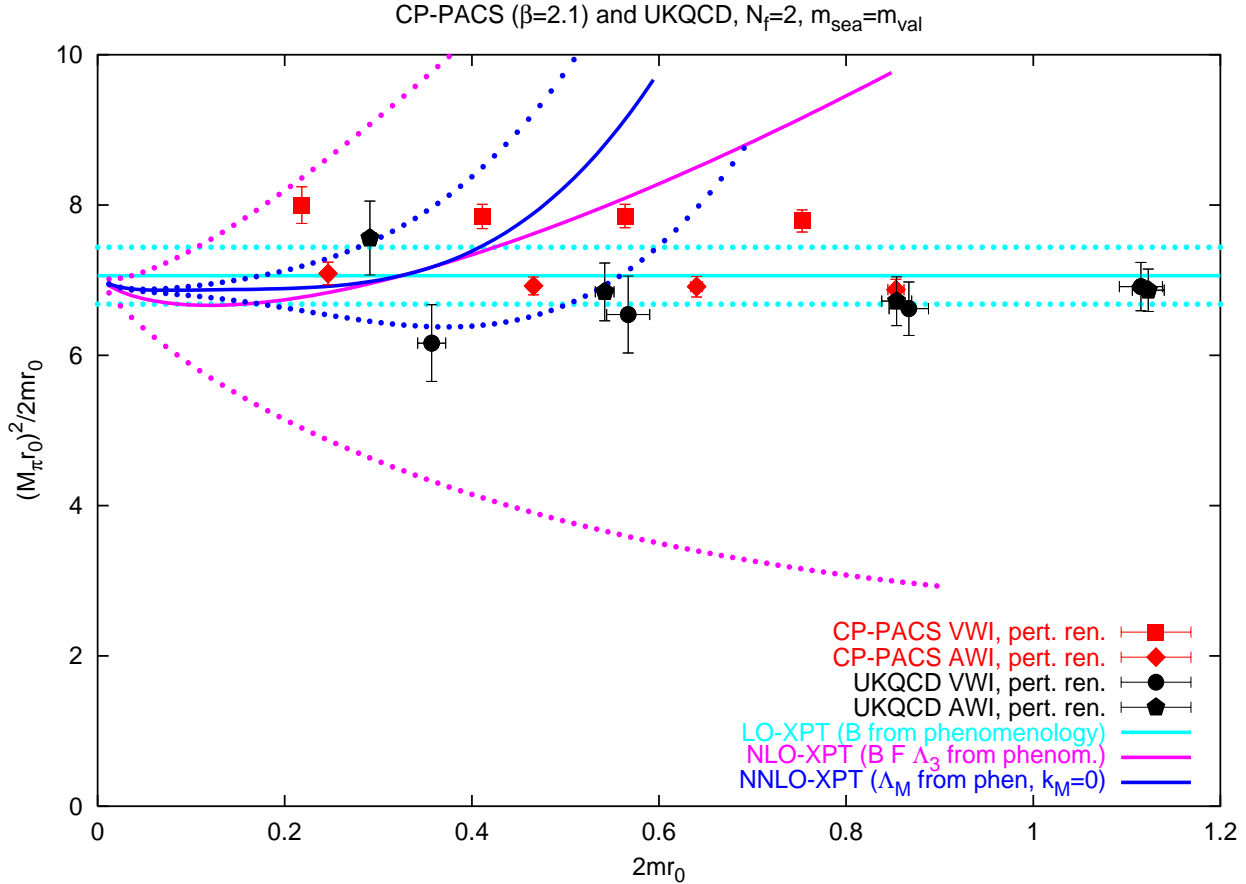,width=16.75cm,height=12cm,angle=0}
\vspace{-4mm}
\caption{\small\sl Same as Fig.\ 2, after dividing by $2mr_0$. Still, lines
represent parameter-free chiral predictions at LO/NLO/NNLO (from light to dark
colour), not fits.}
\end{figure}

Since everything below is about the deviation from a linear relationship, it is
useful to make the curvature optically visible.
This is conveniently done by dividing out a factor $2mr_0$.
The result is displayed in Fig.\ 4 where the predictions from XPT at
LO/NLO/NNLO are included for completeness.
In this representation, the genuine feature of the NLO curve is that it lies
below the LO constant for light pions, but above if the (LO-) pion mass is
larger than $\Lambda_3$.

\begin{figure}
\vspace{-4mm}
\epsfig{file=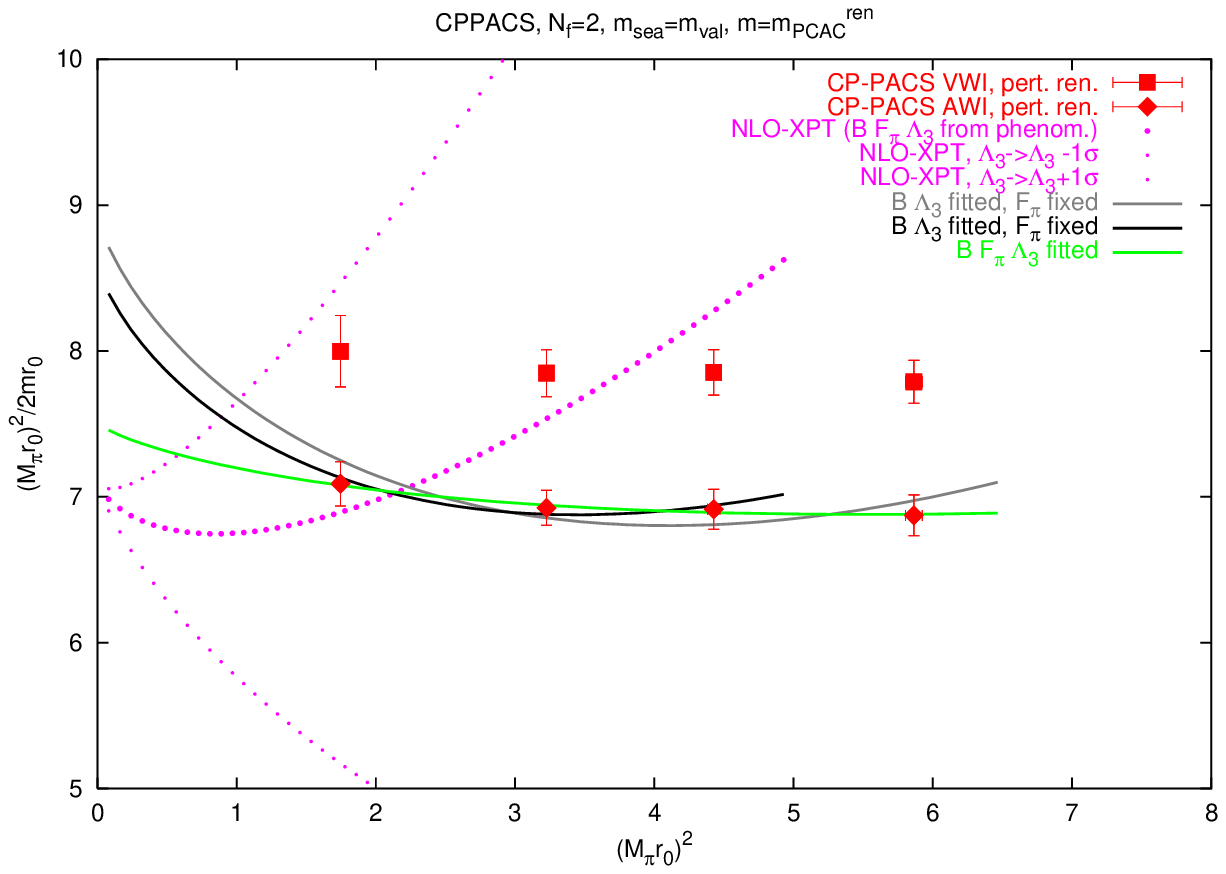,width=16.75cm,height=12cm,angle=0}
\vspace{-2mm}
\\
\epsfig{file=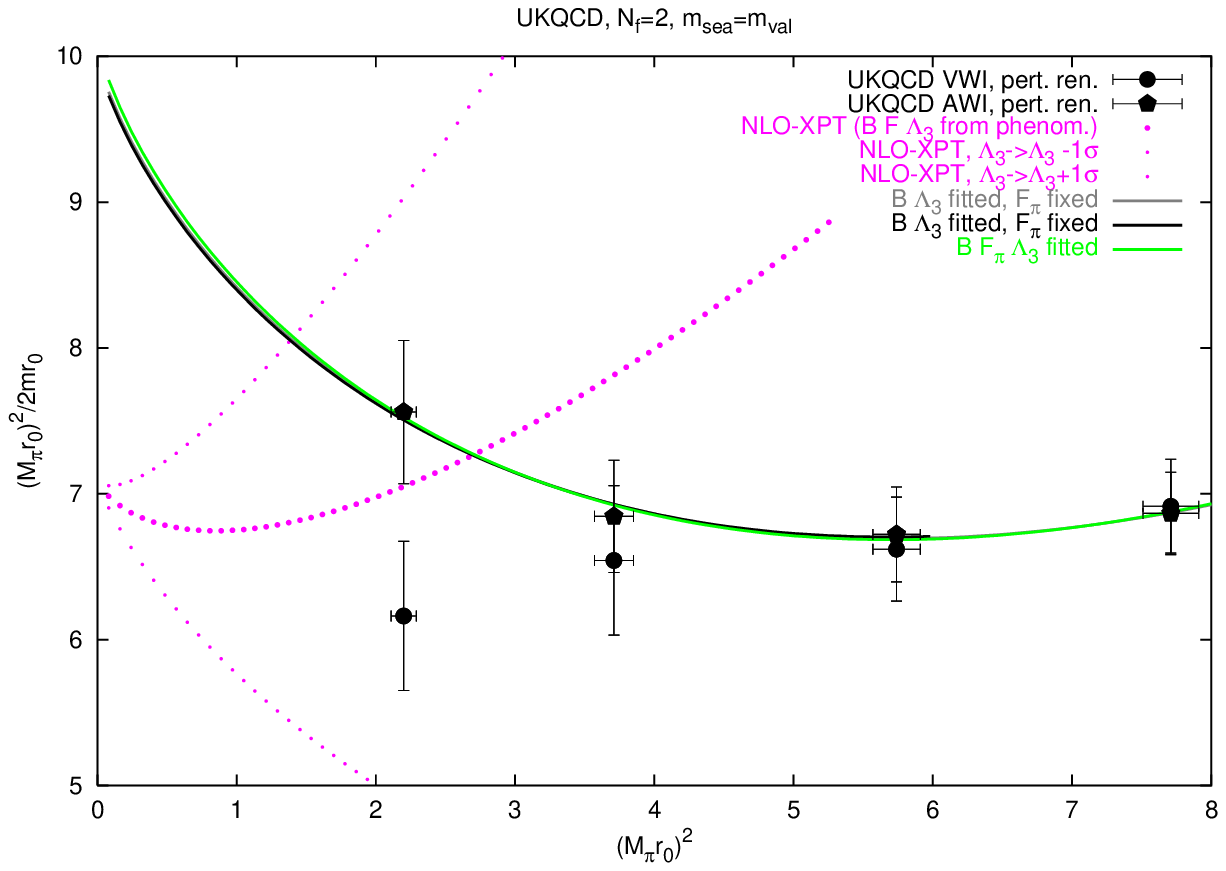,width=16.75cm,height=12cm,angle=0}
\vspace{-4mm}
\caption{\small\sl Fits of the data with the AWI quark mass to the NLO
functional form (together with the allowed band from phenomenological estimate
of $\Lambda_3$). In spite of the bound (\ref{range}) being ignored, the data
fit well to the suggested form (even if one insists on the phenomenological
value of $F_\pi$ in the NLO ansatz (\ref{Mpiult}, \ref{Mpifit})), but with the
current error-bars the LO functional form is still appropriate.}
\vspace{-1.5mm}
\end{figure}

To fit the data, one replaces $2mB$ by $M_\pi^2$, since in this form the data
are plotted against something which is directly measured and the change is of
yet higher order in the chiral counting.
In other words, the statement is that one should use the representation
\beq
{\til M_\pi^2\ovr 2\til m}=
\til B \, \Big\{1-{\til M_\pi^2 \ovr 32\pi^2 \til F_\pi^2}
\log\Big({\til\Lambda_3^2 \ovr \til M_\pi^2}\Big)\Big\}
\label{Mpifit}
\eeq
and adjust the 3 dimensionless parameters $\til B\!\equiv\!Br_0,
\til F\!\equiv\!F_\pi r_0, \til\Lambda_3\!\equiv\!\Lambda_3 r_0$.

\begin{table}
\begin{center}
\begin{tabular}{|l|cccc|cccc|}
\hline
{} & $Br_0$ & $F_\pi r_0$ & $\Lambda_3 r_0$ & $\ch^2/\mr{d.o.f.}$
   & $Br_0$ & $F_\pi r_0$ & $\Lambda_3 r_0$ & $\ch^2/\mr{d.o.f.}$\\
\hline
phenom.\,\cite{Gasser:1983yg, AliKhan:2001tx} & 7.09 & 0.234 & 1.52 & ---
                                              & 7.09 & 0.234 & 1.52 & --- \\
\hline
CP-PACS-a  & 10.1 &(0.234)& 3.32  & 2.63/2  &  8.92 &(0.234)& 3.34  & 1.67/2\\
CP-PACS-b  & 9.65 &(0.234)& 3.01  & 0.23/1  &  8.58 &(0.234)& 3.06  & 0.16/1\\
CP-PACS-c  & 8.30 & 0.577 & 4.16  & 0.08/1  &  7.50 & 0.458 & 3.87  & 0.04/1\\
\hline
CP-PACS-LO & 7.85 &  ---  &  ---  & 0.61/3  &  6.93 &  ---  &  ---  & 0.67/3\\
\hline
UKQCD-a    & 8.05 &(0.234)& 3.21  & 0.73/2  &  10.0 &(0.234)& 3.95  & 0.04/2\\
UKQCD-b    & 7.62 &(0.234)& 2.89  & 0.32/1  &  9.97 &(0.234)& 3.93  & 0.04/1\\
UKQCD-c    & 5.98 &   ?   &   ?   & 0.09/1  &  10.1 & 0.231 & 3.94  & 0.04/1\\
\hline
UKQCD-LO   & 6.67 &  ---  &  ---  & 1.39/3  &  6.90 &  ---  &  ---  & 1.51/3\\
\hline
\end{tabular}
\vspace{-6mm}
\end{center}
\caption{\sl\small Coefficients in the fits of the NLO functional form
$(\ref{Mpifit})$ to the degenerate CP-PACS and UKQCD data with VWI (left) or
AWI (right) definition of the quark masses, after 1-loop renormalization as
detailed in the appendix. Phenomenological values for comparison; constrained
values in brackets. In one case, there is such a shallow minimum that the
fitting algorithm fails to determine it. In general, $\ch^2$ is underestimated,
since the fact that errors are correlated has been ignored. Since the main
uncertainty is systematic, I refrain from quoting statistical errors.}
\vspace{-2mm}
\end{table}

Attempts to fit the data with either the VWI or the AWI definition of the quark
mass to formula (\ref{Mpifit}) are summarized in Tab.\ 1 and -- for the AWI
mass -- illustrated in Fig.\ 5.
Since the parameter $\til F_\pi\!=\!F_\pi r_0\!=\!92.4\,\mr{MeV}\,0.5\,\mr{fm}%
\!\simeq\!0.234$ is known, it is held fixed at its physical value in (a, b) and
fitted only in case (c).
Fit (b) differs from (a) in that the data point with the heaviest quark mass
has been omitted.
From the overall spread
it is clear that there are considerable systematic uncertainties and this is
why I refrain from quoting statistical errors.

It is clear that with the quality of the present data one cannot claim direct
evidence for chiral logs.
Nonetheless, it is worth noticing that fits (a) and (b) indicate that the data
are {\em consistent\/} with the logarithmic form (\ref{Mpifit}) suggested by
NLO chiral perturbation theory.
Moreover, that the fit with $F_\pi$ held fixed at its physical value
does not dramatically change if the heaviest data point is omitted (a$\to$b),
suggests that the estimate (\ref{range}) for the permissible range at NLO level
is indeed reasonable.
What one can learn from Table 1 is that in QCD with $\Nf\!=\!2$ the low-energy
parameter $\Lambda_3$ does not differ by orders of magnitude from the original
estimate by Gasser and Leutwyler \cite{Gasser:1983yg}; all  $\Lambda_3$ values
in Tab.\ 1 lie between the central and the $+1\si$ level in (\ref{l34esti}).
This is important, since it excludes -- at least for $\Nf\!=\!2$ -- an
alternative scenario of the chiral symmetry breaking in which $B$ would not be
the adequate order parameter \cite{GXPTbasis} (for an introduction to this
topic see \cite{Leutwyler:2000hx}).


\section*{Discussion}

The aim of the present note has been to compare the CP-PACS and the UKQCD data
for $M_\pi^2$ as a function of the quark mass to each other and to the
prediction from Chiral Perturbation Theory (XPT).
We have seen that after renormalizing with 1-loop lattice perturbation theory
the two sets are reasonably consistent (maybe with the exception of the CP-PACS
data with the VWI definition of the quark mass).
After dividing out a factor $2mr_0$ to ``zoom in'' on the deviation from a
linear behaviour, both sets show very little curvature.
Postponing the issue whether or not it is permissible to compare
non-continuum-extrapolated lattice data to the prediction from XPT in the
continuum we found that the current size of the error-bars assures
{\em consistency\/} with the logarithmic form (\ref{Mpifit}), albeit one cannot
claim evidence for chiral logs.

The situation might change as more precise data are being released.
In fact, the JLQCD collaboration claims that their data with $\Nf\!=\!2$
non-perturbatively $O(a)$ improved clover fermions and standard (Wilson) glue
are {\em inconsistent\/} with NLO chiral perturbation theory, if they keep
the parameter $F_\pi$ fixed at its physical value \cite{Aoki:2001yr,
BostonHashimoto}.

\bigskip

At this point it is mandatory to think about possible reasons why the data
might not coincide with the chiral prediction.
An incomplete list is the following one:
\begin{enumerate}
\item
They need not, if the pions are too heavy.
Most of the data have been obtained in a regime of quark masses where there is
no guarantee that XPT works.
Phenomenological experience tells us that this expansion works for mesons as
heavy as the physical $K$, i.e.\ for $(M_\pi r_0)^2\!\leq\!1.58$.
I have argued that this condition might be relaxed to (\ref{range}).
Even if this is true, at best 1 point from either set survives, and hence there
is, strictly speaking, no room for trying the NLO-XPT ansatz which has 3
parameters.
\item
Scaling violations, in particular implications of the broken chiral symmetry
might be so severe that
the chiral logs -- even if they exist in the mass range considered -- might
be buried.
To check one would have to perform a continuum extrapolation or attempt a
dynamical simulation with fermions which obey the Ginsparg-Wilson relation.
\item
For unknown reasons (e.g.\ an algorithmic flaw), the data might represent a
partially quenched rather than the fully unquenched situation.
Obviously, this is a rather remote possibility. The point is that the NLO
prediction $M_\pi^2/M^2\!=\!1+\mr{const}M^2\log(M^2/\Lambda_3^2)$ has no
{\em genuine\/} $M^2$ term.
On the other hand, in the partially quenched case a true $M^2$ contribution
exists \cite{Bernard:1993sv, Sharpe:1997by}; hence even perfectly linear
behaviour in a plot analogous to Fig.\ 4 or 5 does not necessarily imply
unreasonable values of the low-energy constants.
\end{enumerate}

\noindent
Finally, I would like to highlight 4 points:

($i$)
So far, we have ignored systematic uncertainties.
Going back to Fig.\ 1 one sees that they are far from being negligible.
What we see exemplifies the standard wisdom that perturbative renormalization
factors which turn out to deviate from 1 by, say, 10\% call for either
repeating the exercise with non-perturbatively determined $Z$-factors or at
much smaller lattice spacing.
In the present context this means that the figures in Tab.\ 1 should not be
taken as lattice determinations of the low-energy constants $F, B, \Lambda_3$.
Nonetheless, even this preliminary attempt shows that $\Lambda_3$ {\em cannot
differ by orders of magnitude\/} from the original estimate by Gasser and
Leutwyler \cite{Gasser:1983yg}.
Moreover, as the data get more precise -- and under the proviso that
consistency is maintained -- lattice determinations of the scales $\Lambda_3$
and $\Lambda_4$ in conjunction with chiral perturbation theory will yield
precise predictions of intrinsically Minkowskian quantities, e.g. the $I=0,2$
pi-pi scattering lengths (see \cite{Leutwyler:2000hx} for the connection).

$(ii)$
For a thorough comparison with chiral perturbation theory, one needs to get
control over the lattice artefacts.
A safe way to do this is to first extrapolate all data to the continuum.
Then one can determine the permissible mass range and extrapolate in a second
step to the chiral limit.
If it turns out that this order cannot be sustained in practice, the chiral
framework may be extended to account for the main discretization effects.
For the case of unimproved fermions this has been done \cite{effbreak}, and it
is clear that this approach could be generalized to improved actions.
The only disadvantage is the added number of counterterms that need to be
fixed from the data.

($iii$)
In the present note the scale is set through $\hat r_0^{-1}$ throughout.
In the CP-PACS studies, $\hat r_0$ has been found to strongly depend on
$\hat m^\mr{VWI}$ or $\hat M_\pi^2$ \cite{AliKhan:2001tx, Aoki:2000kp}.
If $r_0$ itself depends on the sea-quark mass, normalizing all masses through
the measured $\hat r_0$ may not be the best way to compare to XPT; maybe one
should use $\hat r_{0,0}$ (the chirally extrapolated version) instead.

($iv$)
Strictly speaking, the ``phenomenological'' value of $l_3^\mr{r}$ (or
$\Lambda_3$) that we used in comparing the lattice data to the prediction from
XPT is not quite adequate, since the phenomenological determination is with
$m_\mr{s}$ fixed at its physical value, while the lattice studies are
$\Nf\!=\!2$ simulations, in other words here $m_\mr{s}$ is sent to infinity.
The connection to the $\Nf\!=\!3$ GL coefficients $L_i^\mr{r}(\mu)$ is given
by \cite{Gasser:1984gg}
\beq
l_3^\mr{r}=
8\Big(2L_6^\mr{r}-L_4^\mr{r}\Big)+
4\Big(2L_8^\mr{r}-L_5^\mr{r}\Big)-
{1\ovr576\pi^2}
\bigg(\log\Big({\lim_{m_\mr{u,d}\to0}M_\et^2\ovr\mu^2}\Big)+1\bigg)
\label{Nf2ToNf3}
\eeq
[the limit on the r.h.s.\ refers to a situation with $m_\mr{s}$ held fixed at
its physical value], and it is tempting to use it as the starting point of a
little gedanken experiment:
Assume the $s$-quark mass would be such that even when it is doubled the chiral
expansion would not break down.
Eqn.\ (\ref{Nf2ToNf3}) tells us that under these circumstances enhancing
$m_\mr{s}$ by a factor $2$ (and hence roughly doubling $M_\et^2$, too) would
lower $l_3^\mr{r}$ by $0.0001$. Since (\ref{l34esti}) translates into
$l_3^\mr{r}(\mu\!\sim\!M_\rh)=0.0008\pm0.0038$ such a shift would be negligible
compared to the error-bar, and it seems therefore reasonable to assume that
the effect of an added third flavour in the simulations, fixed at the physical
$s$-quark mass (which would fully justify the comparison with $\Nf\!=\!2$ XPT),
would be small compared to the inherent theoretical uncertainties.

Formula (\ref{Nf2ToNf3}) is interesting in yet another respect, since it tells
us that lattice studies which pin down $M_\pi^2$ versus $2m$ (and hence
$\Lambda_3$ or $l_3^\mr{r}$) have a say in another issue.
In the 3-flavour theory there is the famous ``Kaplan Manohar ambiguity'',
i.e.\ the chiral Lagrangian stays invariant under a simultaneous transformation
of the quark masses
\beq
m_\mr{u} \to m_\mr{u}+\la m_\mr{d} m_\mr{s},\;
m_\mr{d} \to m_\mr{d}+\la m_\mr{s} m_\mr{u},\;
m_\mr{s} \to m_\mr{s}+\la m_\mr{u} m_\mr{d},
\eeq
and an appropriate modification of $L_6^\mr{r}, L_7^\mr{r}, L_8^\mr{r}$.
This would, in principle, allow to tune $m_\mr{u}\!=\!0$ and hence provide a
simple and elegant solution to the strong CP problem \cite{Kaplan:1986ru}.
In XPT terminology it is the low-energy combination
$2L_8^\mr{r}(\mu)-L_5^\mr{r}(\mu)$ that decides whether this is a viable
option \cite{Leutwyler:1989pn}.
In the past, XPT has been augmented by theoretical assumptions or model
calculations to exclude $m_\mr{u}\!=\!0$.
More recently it has been proposed to determine the $L_i^\mr{r}$ from
lattice simulations \cite{ShaShoCoKaNe} and important steps in this program
have been taken \cite{CPstrong}.
Formula (\ref{Nf2ToNf3}) tells us that a lattice determination of $l_3^\mr{r}$,
augmented by knowledge about $L_4^\mr{r}$ and $L_6^\mr{r}$, helps to constrain
$m_\mr{u}\!=\!0$.

\bigskip

The present analysis certainly emphasizes the need to compute renormalization
factors non-perturbatively and to perform a continuum extrapolation with
dynamical data.


\subsection*{Acknowledgements}
\vspace{-4pt}
This work has taken its origin in Seattle, during the INT program ``Lattice QCD
and Hadron Phenomenology''. Discussions with Maarten Golterman and Rainer
Sommer are gratefully acknowledged, as well as useful correspondence with Derek
Hepburn and Akira Ukawa. I am indebted to Heiri Leutwyler for
providing me with phenomenological estimates of the NNLO constant $k_M$ and to
Stefano Capitani for a check of the generic formulas in appendices A,B.


\clearpage

\section*{Appendix A: Renormalization of the VWI quark mass}

Setup with $m^\mr{VWI}a=\log(1+{1\ovr2}({1\ovr\ka}-{1\ovr\ka_\mr{c}}))
\simeq{1\ovr2}({1\ovr\ka}-{1\ovr\ka_\mr{c}})$ and
$u_0=P^{1/4}=({1\ovr3}\<\mr{Tr}U_\Box\>)^{1/4}$ \cite{Lepage:1992xa}:
\bea
m_\MSbar^\mr{VWI}(\mu)&=&Z_m(\mu a) \Big(1+b_m am\Big) m
\label{A01}
\\
m_\MSbar^\mr{VWI}(\mu)&=&\til Z_m(\mu a) \Big(1+\til b_m {am \ovr u_0}\Big)
 {m \ovr u_0}
\label{A02}
\eea

\noindent
From \cite{Gupta:1996sa}, naive versus tadpole-improved, with $g^2=6/\be$ and
$\til g^2=\til g_\MSbar^2(2\GeV)$:
\bea
Z_m(\mu a)     &=&1+{g^2\ovr4\pi}
\Big( {z_m\ovr3\pi}-{1\ovr\pi}\log(a^2\mu^2) \Big)
\label{A03}
\\
\til Z_m(\mu a)&=&1+{\til g^2\ovr4\pi}
\Big( {\til z_m\ovr3\pi}-{1\ovr\pi}\log(a^2\mu^2) \Big)
\label{A04}
\eea

\noindent
From \cite{Aoki:1998ar} (key to orig.\ literature: their [21-28]), for generic
actions:
\beq
{1\ovr g_\MSbar^2(\mu)}={1\ovr g^2}+
d_g+d_f \Nf + {11-2\Nf/3\ovr16\pi^2}\log(a^2\mu^2)
\label{A05}
\eeq

Wilson/Clover: $d_g=-0.4682, d_f=0.0314917$ (for $c_\mr{SW}=1$) and
$P=1-1/3\cdot g^2$
\bea
{1\ovr g_\MSbar^2(\mu)}&=&{1\ovr g^2}
-0.4682+0.0314917\Nf + {11-2\Nf/3\ovr16\pi^2}\log(a^2\mu^2)
\label{A06}
\\
{1\ovr \til g_\MSbar^2(\mu)}&=&{P\ovr g^2}
-0.1349+0.0314917\Nf + {11-2\Nf/3\ovr16\pi^2}\log(a^2\mu^2)
\label{A07}
\\
z_m &=& 12.953+7.738c_\mr{SW}-1.380c_\mr{SW}^2 
\label{A08}
\\
\til z_m &=& z_m-\pi^2 \;=\;
\left\{
\begin{tabular}{rc}
13.0&($c_\mr{SW}\simeq2$)\\
9.44&($c_\mr{SW}=1$)
\end{tabular}
\right.
\label{A09}
\eea

Iwasaki/Clover: $d_g=0.1000, d_f=0.0314917$ (for $c_\mr{SW}=1$) and
$P=1-0.1402 g^2$,

$R=1-0.2689 g^2$, thus
$1=P+0.1402 g^2 = R+0.2689 g^2 = 3.648P-8\cdot0.331R-0.2006 g^2$
\bea
{1\ovr g_\MSbar^2(\mu)}&=&{1\ovr g^2}
+0.1000+0.0314917\Nf + {11-2\Nf/3\ovr16\pi^2}\log(a^2\mu^2)
\label{A10}
\\
{1\ovr \til g_\MSbar^2(\mu)}&=&{P\ovr g^2}+0.2402
+0.0314917\Nf + {11-2\Nf/3\ovr16\pi^2}\log(a^2\mu^2)
\label{A11}
\\
{1\ovr \til g_\MSbar^2(\mu)}&=&{R\ovr g^2}+0.3689
+0.0314917\Nf + {11-2\Nf/3\ovr16\pi^2}\log(a^2\mu^2)
\label{A12}
\\
{1\ovr \til g_\MSbar^2(\mu)}&=&{3.648P-2.648R \ovr g^2}-0.1006
+0.0314917\Nf + {11-2\Nf/3\ovr16\pi^2}\log(a^2\mu^2)
\label{A13}
\\
z_m &=& 4.858+5.301c_\mr{SW}-1.267c_\mr{SW}^2 
\label{A14}
\\
\til z_m&=&z_m-0.4206\pi^2 \;=\;
\left\{
\begin{tabular}{rc}
5.76&($c_\mr{SW}\!=\!1.47$)\\
4.74&($c_\mr{SW}=1$)
\end{tabular}
\right.
\label{A15}
\eea

\noindent
In \cite{Aoki:1998qd} one finds:
\bea
\mr{Wilson/Clover}: \qquad
b_m &=& -1/2 -0.09623 g^2 +O(g^4)
\label{A16}
\\
\til b_m &=& b_m(1-1/12\cdot\til g^2) \;\simeq\; -1/2-0.05456 \til g^2
\label{A17}
\\
\mr{Iwasaki/Clover}:\qquad
b_m &=& -1/2 -0.0509 g^2 +O(g^4)
\label{A18}
\\
\til b_m &=& b_m(1-0.03505\til g^2) \;\simeq\; -1/2-0.0334 \til g^2
\label{A19}
\eea


\section*{Appendix B: Renormalization of the AWI quark mass}

Setup with
$m^\mr{AWI, imp}a={1\ovr2}\<\pad_\mu A_\mu^\mr{a, imp}(x) O^\mr{a}(0)\>/
\<P^\mr{a}(x) O(0)\>$ and $u_0=P^{1/4}=({1\ovr3}\<\mr{Tr}U_\Box\>)^{1/4}$
\cite{Lepage:1992xa}:
\bea
m_\MSbar^\mr{AWI}(\mu)&=&{Z_A\ovr Z_P(\mu a)} \;
 {1+b_A am \ovr 1+b_P am} \; m^\mr{AWI,imp}
\label{B01}
\\
m_\MSbar^\mr{AWI}(\mu)&=&{\til Z_A\ovr \til Z_P(\mu a)} \;
 {1+\til b_A am/u_0 \ovr 1+\til b_P am/u_0} \; m^\mr{AWI,imp}
\label{B02}
\eea

\noindent
From \cite{Gupta:1996sa}, naive versus tadpole-improved, with $g^2=6/\be$ and
$\til g^2=\til g_\MSbar^2(2\GeV)$:
\bea
Z_A              &=&1+{g^2\ovr4\pi} \; {z_A\ovr3\pi}\;,\qquad
Z_P(\mu a)      \;=\; 1+{g^2\ovr4\pi}
\Big( {z_P\ovr3\pi}+{1\ovr\pi}\log(a^2\mu^2) \Big)
\label{B03}
\\
\til Z_A         &=&1+{\til g^2\ovr4\pi} \; {\til z_A\ovr3\pi}\;,\qquad
\til Z_P(\mu a) \;=\; 1+{\til g^2\ovr4\pi}
\Big( {\til z_P\ovr3\pi}+{1\ovr\pi}\log(a^2\mu^2) \Big)
\label{B04}
\eea

\noindent
From \cite{Taniguchi:1998pf}, for generic actions:

Wilson/Clover:
\bea
z_A &=& -15.797-0.248c_\mr{SW}+2.251c_\mr{SW}^2 
\label{B05}
\\
\til z_A &=& z_A+\pi^2 \;=\;
\left\{
\begin{tabular}{rc}
2.581&($c_\mr{SW}\simeq2$)\\
-3.924&($c_\mr{SW}=1$)
\end{tabular}
\right.
\label{B06}
\\
z_P &=& -22.596+2.249c_\mr{SW}-2.036c_\mr{SW}^2 
\label{B07}
\\
\til z_P &=& z_P+\pi^2 \;=\;
\left\{
\begin{tabular}{rc}
-16.372&($c_\mr{SW}\simeq2$)\\
-12.513&($c_\mr{SW}=1$)
\end{tabular}
\right.
\label{B08}
\eea

Iwasaki/Clover:
\bea
z_A &=& -8.192-0.125c_\mr{SW}+1.610c_\mr{SW}^2 
\label{B09}
\\
\til z_A &=& z_A+0.4206\pi^2 \;=\;
\left\{
\begin{tabular}{rc}
-0.746&($c_\mr{SW}\!=\!1.47$)\\
-2.556&($c_\mr{SW}=1$)
\end{tabular}
\right.
\label{B10}
\\
z_P &=& -10.673+1.601c_\mr{SW}-1.281c_\mr{SW}^2 
\label{B11}
\\
\til z_P &=& z_P+0.4206\pi^2 \;=\;
\left\{
\begin{tabular}{rc}
-6.936&($c_\mr{SW}\!=\!1.47$)\\
-6.202&($c_\mr{SW}=1$)
\end{tabular}
\right.
\label{B12}
\eea

\noindent
In \cite{Aoki:1998qd} one finds:
\bea
\mr{Wilson/Clover}: \qquad
b_A &=& 1 +0.15219(5) g^2 +O(g^4)
\label{B13}
\\
\til b_A &=& b_A(1-1/12\cdot\til g^2) \;\simeq\; 1+0.06886(5) \til g^2
\label{B14}
\\
b_P &=& 1 +0.15312(3) g^2 +O(g^4)
\label{B15}
\\
\til b_P &=& b_P(1-1/12\cdot\til g^2) \;\simeq\; 1+0.06979(3) \til g^2
\label{B16}
\\
\mr{Iwasaki/Clover}: \qquad
b_A &=& 1 +0.0733(5) g^2 +O(g^4)
\label{B17}
\\
\til b_A &=& b_A(1-0.03505\til g^2) \;\simeq\; 1+0.0383(5) \til g^2
\label{B18}
\\
b_P &=& 1 +0.0744(12) g^2 +O(g^4)
\label{B19}
\\
\til b_P &=& b_P(1-0.03505\til g^2) \;\simeq\; 1+0.0394(12) \til g^2
\label{B20}
\eea
In all cases the relationship between $z_X$ and $\til z_X$ as well as $b_X$ and
$\til b_X$ ($X\in\{m,A,P\}$) is mine.


\section*{Appendix C: Implementation with tadpole resummation}

The renormalization with $u_0$ defined via the plaquette is traced in Tab.\ 2
for the UKQCD data.
Alternatively, one could define $u_0$ as $1/(8\ka_\mr{c})$, but this would
bring different perturbative coefficients than those listed in App.\ A/B.
Notice the effect of the tadpole resummation, i.e.\ the difference of the
coupling $g_\MSbar^2$ computed via (\ref{A06}) and $\til g_\MSbar^2$ via
(\ref{A07}).
The statistical uncertainty for the VWI quark mass is larger than that for the
AWI definition due to a limited accuracy of $\ka_\mr{c}$, which is defined in
a partially quenched sense \cite{Allton:2001sk}.

\begin{table}[!h]
\begin{center}
\begin{tabular}{|l|cccc|}
\hline
{\bf UKQCD} & (5.20,0.1355) & (5.20,0.1350) & (5.26,0.1345) & (5.29,0.1340)
\\
\hline 
$\hat M_\pi=M_\pi a$ \hfill\cite{Allton:2001sk} &
 0.294(4) & 0.405(4) & 0.509(2) & 0.577(2)
\\
$\hat r_0=r_0/a$ \hfill\cite{Allton:2001sk} &
 5.041(40) & 4.754(40) & 4.708(52) & 4.813(45)
\\
$(M_\pi r_0)^2$ &
 2.20(09) & 3.71(14) & 5.74(17) & 7.71(20)
\\
\hline 
$\ka_\mr{c}$ \hfill\cite{Allton:2001sk} &
 0.13645(3) & 0.13663(5) & 0.13709(3) & 0.13730(3)
\\
$\hat m^\mr{VWI}\equiv{1\ovr2}({1\ovr\ka}-{1\ovr\ka_\mr{c}})$ &
 0.0257(08) & 0.0442(13) & 0.0702(08) & 0.0897(08)
\\
$\hat m^\mr{AWI, imp}$ \hfill\cite{Allton:2001sk} &
 0.0231(3) & 0.0462(3) & 0.0742(3) & 0.0952(3)
\\
$P$ \hfill\cite{Booth:2001qp} &
 0.536294(9) & 0.533676(9) & 0.539732(9) & 0.542410(9)
\\
$(a\,2\GeV)^2\equiv(5.06773/\hat r_0)^2$ &
 1.011(16) & 1.136(19) & 1.159(26) & 1.109(21)
\\
$g_\MSbar^2(2\GeV)$ \hfill(\ref{A06}) &
 2.1640(45) & 2.1309(46) & 2.0813(58) & 2.0714(49)
\\
$\til g_\MSbar^2(2\GeV)$ \hfill(\ref{A07}) &
 2.541(6) & 2.510(6) & 2.437(8) & 2.423(7)
\\
$\til b_m$ \hfill(\ref{A17}) &
 -0.6387(3) & -0.6369(4) & -0.6330(4) & -0.6322(4)
\\
$\til b_A$ \hfill(\ref{B14}) &
 1.1750(4) & 1.1728(4) & 1.1678(5) & 1.1669(5)
\\
$\til b_P$ \hfill(\ref{B16}) &
 1.1773(4) & 1.1752(4) & 1.1701(6) & 1.1691(5)
\\
\hline 
$c_\mr{SW}$ \hfill\cite{Allton:2001sk} &
 \multicolumn{2}{c}{2.0171} & 1.9497 & 1.9192
\\
$\til z_m$ \hfill(\ref{A08}, \ref{A09}) &
 \multicolumn{2}{c}{13.0769} & 12.9243 & 12.8512
\\
$\til z_A$ \hfill(\ref{B05}, \ref{B06}) &
 \multicolumn{2}{c}{2.73099} & 2.14587 & 1.88782
\\
$\til z_P$ \hfill(\ref{B07}, \ref{B08}) &
 \multicolumn{2}{c}{-16.4738} & -16.081 & -15.9094
\\
$\til Z_m(a\,2\GeV)$ \hfill(\ref{A04}) &
 1.2799(17) & 1.2690(18) & 1.2569(22) & 1.2566(18)
\\
$\til Z_A$ \hfill(\ref{B04}) &
 1.0586(1) & 1.0579(1) & 1.0442(1) & 1.0386(1)
\\
$\til Z_P(a\,2\GeV)$ \hfill(\ref{B04}) &
 0.6472(1) & 0.6590(2) & 0.6781(3) & 0.6808(3)
\\
\hline 
$2m_\MSbar^\mr{VWI}(2\GeV) r_0$ \hfill(\ref{A02}) &
 0.380(16) & 0.603(25) & 0.919(23) & 1.181(24)
\\
$2m_\MSbar^\mr{AWI}(2\GeV) r_0$ \hfill(\ref{B02}) &
 0.381(09) & 0.705(13) & 1.076(19) & 1.398(21)
\\
\hline 
$c_\mr{SW}$ &
 \multicolumn{4}{c|}{1}
\\
$\til z_m, \til z_A, \til z_P$ \hfill(\ref{A09}, \ref{B06}, \ref{B08}) &
 \multicolumn{4}{c|}{9.4414 , -3.9244 , -12.5134}
\\
$\til Z_m(a\,2\GeV)$ \hfill(\ref{A04}) &
 1.2019(15) & 1.192(16) & 1.1852(20) & 1.1869(17)
\\
$\til Z_A$ \hfill(\ref{B04}) &
 0.9158(2) & 0.9168(2) & 0.9192(3) & 0.9197(2)
\\
$\til Z_P(a\,2\GeV)$ \hfill(\ref{B04}) &
 0.7322(4) & 0.7429(4) & 0.7516(5) & 0.7503(5)
\\
\hline 
$2m_\MSbar^\mr{VWI}(2\GeV) r_0$ \hfill(\ref{A02}) &
 0.357(15) & 0.567(23) & 0.867(21) & 1.115(23)
\\
$2m_\MSbar^\mr{AWI}(2\GeV) r_0$ \hfill(\ref{B02}) &
 0.291(07) & 0.542(10) & 0.854(16) & 1.123(17)
\\
\hline 
\end{tabular}
\vspace{-4mm}
\end{center}
\caption{\sl\small
Step-by-step renormalization of UKQCD quark masses with $u_0$ defined via the
plaquette. Error bars include statistical errors only, naive error propagation
throughout. Note the disparity of $g_\MSbar^2$ defined via (\ref{A06}) and
that via (\ref{A07}). It makes a difference whether $c_\mr{SW}$ as used in the
simulations is plugged in or $c_\mr{SW}=1$ which is a consistent choice at the
order we are interested in.}
\end{table}

For the CP-PACS data, both the plaquette $P$ and the rectangle $R$ are
published \cite{AliKhan:2001tx}, and this gives, in principle, the option to
define the tadpole resummation (besides the usual option $1/(8\ka_\mr{c})$) via
(\ref{A12}) or (\ref{A13}), where the latter choice reflects the specific
combination used in the action.
All these options would, however, imply different perturbative coefficients
than those listed in App.\ A/B, i.e.\ we restrict ourselves to the choice
(\ref{A11}) along with $u_0\equiv P^{1/4}$.
It is interesting to see that the CP-PACS RG improved action achieves agreement
of $g_\MSbar^2$ with the ``standard'' $\til g_\MSbar^2$ (via (\ref{A11})).

In the argument of the logarithm that converts to the $\MSbar$ scheme, the
lattice spacing is multiplied with a physical scale, here $2\GeV$.
This means that $a$ must be assigned a physical value, as well.
To compare like with like this is done via the measured $r_0$ in both sets,
assuming $r_0=0.5\fm$ in physical units.
As a consequence, our $Z$ factors for the CP-PACS data depend slightly on
the quark mass, even though we work in a mass-independent scheme.

\begin{table}[!h]
\begin{center}
\begin{tabular}{|l|cccc|}
\hline
{\bf CP-PACS} & (2.10,0.1382) & (2.10,0.1374) & (2.10,0.1367) & (2.10,0.1357)
\\
\hline 
$\hat M_\pi=M_\pi a$ \hfill\cite{AliKhan:2001tx} &
 0.29459(85) & 0.42401(46) & 0.51671(67) & 0.63010(61)
\\
$\hat r_0=r_0/a$ \hfill\cite{AliKhan:2001tx} &
 4.485(12) & 4.236(14) & 4.072(15) & 3.843(16)
\\
$(M_\pi r_0)^2$ &
 1.746(19) & 3.226(28) & 4.427(44) & 5.864(60)
\\
\hline 
$\ka_\mr{c}$ \hfill\cite{AliKhan:2001tx} &
 \multicolumn{4}{c|}{0.138984(13)}
\\
$\hat m^\mr{VWI}\equiv{1\ovr2}({1\ovr\ka}-{1\ovr\ka_\mr{c}})$ &
 0.02041(34) & 0.04147(34) & 0.06011(34) & 0.08706(34)
\\
$\hat m^\mr{AWI, imp}$ \hfill\cite{AliKhan:2001tx} &
 0.02613(18) & 0.05267(22) & 0.07564(38) & 0.10748(51)
\\
$P$ \hfill\cite{AliKhan:2001tx} &
 0.6010819(84) & 0.6000552(67) & 0.5992023(76) & 0.5980283(76)$\!$
\\
$R$ \hfill\cite{AliKhan:2001tx} &
 0.366883(13) & 0.365297(10) & 0.363979(12) & 0.362139(12)
\\
$u_0 \equiv P^{1/4}$ &
 0.880508(3) & 0.880132(2) & 0.879819(3) & 0.879388(3)
\\
$(a\,2\GeV)^2\equiv(5.06773/\hat r_0)^2$ &
 1.277(07) & 1.431(09) & 1.549(11) & 1.739(14)
\\
$g_\MSbar^2(2\GeV)$ \hfill(\ref{A10}) &
 1.8941(12) & 1.8694(14) & 1.8526(15) & 1.8286(17)
\\
$\til g_\MSbar^2(2\GeV)$ \hfill(\ref{A11}) &
 1.8920(12) & 1.8686(14) & 1.8529(15) & 1.8303(17)
\\
$\til g_\MSbar^2(2\GeV)$ \hfill(\ref{A12}) &
 1.7383(10) & 1.7191(12) & 1.7063(13) & 1.6877(14)
\\
$\til g_\MSbar^2(2\GeV)$ \hfill(\ref{A13}) &
 2.4705(20) & 2.4276(24) & 2.3986(26) & 2.3574(28)
\\
$\til b_m$ \hfill(\ref{A19}) &
 -0.56315(4) & -0.56236(5) & -0.56184(5) & -0.56108(6)
\\
$\til b_A$ \hfill(\ref{B18}) &
 1.07237(4) & 1.07147(5) & 1.07087(6) & 1.07001(7)
\\
$\til b_P$ \hfill(\ref{B20}) &
 1.07445(5) & 1.07353(6) & 1.07291(6) & 1.07202(7)
\\
\hline 
$c_\mr{SW}$ \hfill\cite{AliKhan:2001tx} &
 \multicolumn{4}{c|}{1.47}
\\
$\til z_m, \til z_A, \til z_P$ \hfill(\ref{A15}, \ref{B10}, \ref{B12}) &
 \multicolumn{4}{c|}{5.76145 , -0.745545 , -6.93649}
\\
$\til Z_m(a\,2\GeV)$ \hfill(\ref{A04}) &
 1.08033(31) & 1.07393(37) & 1.06960(40) & 1.06338(44)
\\
$\til Z_A$ \hfill(\ref{B04}) &
 0.98809(1) & 0.98824(1) & 0.98834(1) & 0.98848(1)
\\
$\til Z_P(a\,2\GeV)$ \hfill(\ref{B04}) &
 0.90090(19) & 0.90753(24) & 0.91201(27) & 0.91846(31)
\\
\hline 
$2m_\MSbar^\mr{VWI}(2\GeV) r_0$ \hfill(\ref{A02}) &
 0.2217(44) & 0.4174(50) & 0.5723(57) & 0.7642(66)
\\
$2m_\MSbar^\mr{AWI}(2\GeV) r_0$ \hfill(\ref{B02}) &
 0.2571(27) & 0.4859(42) & 0.6675(65) & 0.8889(89)
\\
\hline 
$c_\mr{SW}$ &
 \multicolumn{4}{c|}{1}
\\
$\til z_m, \til z_A, \til z_P$ \hfill(\ref{A15}, \ref{B10}, \ref{B12})&
 \multicolumn{4}{c|}{4.74084 , -2.55584 , -6.20184}
\\
$\til Z_m(a\,2\GeV)$ \hfill(\ref{A04}) &
 1.06403(30) & 1.05783(36) & 1.05363(39) & 1.04761(43)
\\
$\til Z_A$ \hfill(\ref{B04}) &
 0.95917(3) & 0.95968(3) & 0.96001(3) & 0.96050(4)
\\
$\til Z_P(a\,2\GeV)$ \hfill(\ref{B04}) &
 0.91263(20) & 0.91912(25) & 0.92351(28) & 0.92981(32)
\\
\hline 
$2m_\MSbar^\mr{VWI}(2\GeV) r_0$ \hfill(\ref{A02}) &
 0.2183(43) & 0.4111(49) & 0.5637(56) & 0.7529(65)
\\
$2m_\MSbar^\mr{AWI}(2\GeV) r_0$ \hfill(\ref{B02}) &
 0.2463(26) & 0.4659(40) & 0.6403(63) & 0.8532(86)
\\
\hline 
\end{tabular}
\vspace{-4mm}
\end{center}
\caption{\sl\small
Step-by-step renormalization of CP-PACS quark masses with $u_0$ defined via the
plaquette. Error bars include statistical errors only, naive error propagation
throughout. Note the similarity of $g_\MSbar^2$ defined via (\ref{A10}) and
that via (\ref{A11}). There is little difference whether $c_\mr{SW}$ as used
in the simulations is plugged in or $c_\mr{SW}=1$ which is a consistent choice
at the order we are interested in.}
\end{table}

\clearpage


\end{document}